\documentclass[12pt]{article}

\newcommand{\bl}{\mbox{\boldmath$l$}}
\newcommand{\pbpi}{\,^+\!\mbox{\boldmath$\pi$}}
\newcommand{\mbpi}{\,^-\!\mbox{\boldmath$\pi$}}
\newcommand{\pmbpi}{\,^{\pm}\!\mbox{\boldmath$\pi$}}
\renewcommand{\d}{{\rm d}}
\newcommand{\bpi}{\mbox{\boldmath$\pi$}}

\begin{document}

\title{Discrete quantum gravity in the framework of Regge calculus formalism}
\author{V.M. Khatsymovsky \\
 {\em Budker Institute of Nuclear Physics} \\ {\em
 Novosibirsk,
 630090,
 Russia}
\\ {\em E-mail address: khatsym@inp.nsk.su}}
\date{}
\maketitle
\begin{abstract}
An approach to the discrete quantum gravity based on the Regge calculus is discussed
which was developed in a number of our papers. Regge calculus is general relativity
for the subclass of general Riemannian manifolds called piecewise flat ones. Regge
calculus deals with the discrete set of variables, triangulation lengths, and contains
continuous general relativity as a particular limiting case when the lengths tend to
zero. In our approach the quantum length expectations are nonzero and of the order of
Plank scale $10^{-33}cm$. This means the discrete spacetime structure on these scales.
\end{abstract}

PACS numbers: 04.60.-m Quantum gravity

\newpage

{\bf 1.Introduction}

\medskip

An interest to formulation of general relativity (GR) in a discrete form is provided,
not in the last place, by complexity of the theory. In classical aspect, rewriting the
essentially nonlinear equations of the theory, Einstein equations, in terms of a
discrete set of physical quantities, that is, discretising them simplifies using
numerical methods for solving these equations. In quantum aspect, discretisation might
be introduced, as in any other field theory, in order to regularise the originally
divergent expressions. However, in the case of GR the following two distinctive
features arise. First, according to the standard classification, GR is
nonrenormalisable theory, therefore dependence of the result on the specific way of
the regularisation cannot be removed by the renormalisation. Consequently, here
discretisation must not only be a mathematical approximation like finite-difference
one of the originally continuum theory but present some realizable physics specifying
the form of the theory at small distances. Second, the covariance with respect to the
arbitrary coordinate transformatioms is specific for GR, and this property is badly
consistent with quantum theory in which the time plays the special role. To avoid this
difficulty, one might try to formulate GR in the explicitly coordinateless form.

In the framework of Regge calculus suggested in 1961 \cite{Regge} the exact GR is
considered on the particular case of general Riemannian spacetime, the so-called
piecewise flat manifolds which are flat everywhere with exception of the subset of
points of zero measure. Any such spacetime can be represented as collection of a
number of the flat 4-dimensional {\it simplices}(tetrahedrons). In the $n$-dimensional
case the $n$-dimensional simplices $\sigma^n$ are considered. The $n$-dimensional
simplex $\sigma^n$ contains the $n+1$ vertices each of which being connected by the
edges with the other $n$ vertices. All the geometrical characteristics of the
$n$-simplex are uniquely defined by the (freely chosen) lengths of the $n{{\textstyle
n+1}\over {\textstyle 2}}$ edges of it. Geometry of the Regge spacetime is defined by
the freely chosen lengths of all edges (or 1-simplices). The linklengths of the two
$n$-simplices sharing some $(n-1)$-simplex as their common face should coincide on
this face. As for all the $n$-simplices containing some $(n-2)$-simplex as
$(n-2)$-dimensional face this manifold cannot be embedded into the flat
$n$-dimensional spacetime if linklengths are freely chosen since the sum of
hyperdihedral angles at $(n-2)$-face in all the $n$-simplices meeting at this face is
$2\pi$ - $\alpha$ where the so-called angle defect $\alpha$ does not generally vanish.
As a result of the parallel transport of a vector along a closed contour contained in
the given $n$-simplices and enclosing the given $(n-2)$-simplex, the vector is rotated
by the angle $\alpha$. This corresponds to a $\delta$-function-like curvature
distribution with support on $(n-2)$-simplices proportional to the angle defects on
these simplices. The action for the 4-dimensional Regge gravity is proportional to
\begin{equation}                                                                    
\label{S-Regge} \sum_{\sigma^2}{\alpha_{\sigma^2}|\sigma^2|},
\end{equation}

\noindent where $|\sigma^2|$ is the area of a triangle (the 2-simplex) $\sigma^2$,
$\alpha_{\sigma^2}$ is the angle defect on this triangle, and summation run over all
the 2-simplices $\sigma^2$. It is shown in the work \cite{Fried} that the action
(\ref{S-Regge}) can be obtained from the expression
\begin{equation}                                                                    
\label{S-Einst} {1\over 2}\int{R\sqrt{g}{\rm d}^4x},
\end{equation}

\noindent to which the Einstein action is proportional when passing to the
$\delta$-function limit of the curvature $R$ distribution. Thus Regge calculus is GR
in which some degrees of freedom are "frozen", that is, the so-called minisuperspace
theory for GR. Thereby the first of above mentioned requirements is satisfied, namely,
Regge manifold is a particular (although somewhat singular) case of general Riemannian
manifold. Besides that, the mutual location of vertices (the 0-dimensional simplices)
and thereby the geometry is uniquely fixed by the values of invariant lengths of the
edges (the 1-simplices $\sigma^1$) which therefore play the role of the field
variables. Thus the second requirement of the coordinateless description is also
fullfilled.

Although Regge calculus is only some subset in the configuration superspace of GR,
this subset is dense in this superspace. That is, each nonsingular Riemannian manifold
can be approximated with arbitrarily high accuracy by an appropriately chosen Regge
manifold. To construct such the Regge manifold one divides, for example, the
Riemannian manifold into the sufficiently small regions topologically equivalent to
the simplices $\sigma^4$, the edges of these being geodesics. As the piecewise-flat
manifold of interest the manifold of such type can be taken possessing the same scheme
(topology) of connection of the different vertices by the edges and the linklengths
being geodesic lengths. It is shown in the paper \cite{Fein} that the Einstein action
(\ref{S-Einst}) follows as limiting case of the Regge action (\ref{S-Regge}) for such
the approximating spaces if a typical edge length for triangulation length) tends to
zero. A more general statement \cite{Cheeger} is that the so-called {\it
Lipshitz-Killing curvatures} converge to their continuum counterparts {\it in the
sense of measures} if the decomposition into the 4-simplices becomes finer and finer,
that is, it is integrals of the considered values over the spacetime regions which
converge (to the integrals of the continuum counterparts). The volume of a spacetime
region, contribution of the region to the Einstein action and to the Gauss-Bonnet
topological term are examples of such the integrals.

Regge calculus possesses exact discrete analogs of many quantities which can be
defined in the continuum GR. The first example are the Einstein equations whose
discrete analog was derived by Regge upon varying the action (\ref{S-Regge}) over the
linklengths. It turns out that the variation of $\alpha_{\sigma^2}$ does not
contribute into (\ref{S-Regge}), and the equation obtained by variation of a
particular edge $\sigma^1$ takes the form
\begin{equation}\label{Reg-equation}                                                
\sum_{\sigma^2\supset\sigma^1}{\alpha_{\sigma^2}\cot\vartheta (\sigma^1,\sigma^2)} =
0.
\end{equation}

\noindent Here $\vartheta (\sigma^1,\sigma^2)$ is the angle in the triangle $\sigma^2$
opposite to the edge $\sigma^1$ while summation is over all the triangles sharing
$\sigma^1$. Evidently, the discrete coordinateless formulation in terms of physical
quantities (lengths) is an ideal means for numerical simulations, and originally Regge
calculus was just used for numerical analysis of the Einstein equations \cite{Wong}.

However Regge calculus is of the maximal interest if applied to quantum gravity. In
this aspect the main problem is in constructing the Hamiltonian formalism analogous to
the Arnowitt-Deser-Misner construction in the continuum GR \cite{ADM}. In accordance
with their result, the GR Lagrangian can be reduced to the form
\begin{equation}                                                                    
L = \sum_A{p_A\dot{q}_A} - \sum_{\alpha}{\lambda_{\alpha}\Phi_{\alpha}(p,q)}
\end{equation}

\noindent with the canonical variables $p_A$, $q_A$ and yet another variables
$\lambda_{\alpha}$ playing the role of the Lagrange multipliers whose dynamics is not
fixed by the equations of motion. Thus, GR is a theory described by the set of
constraints $\Phi_{\alpha}(p,q)$ = 0 and zero Hamiltonian. In the case of
coordinateless Regge calculus theory we need to particularly return to the coordinate
description but with respect to only time coordinate $t$, the discrete field
distribution (here lengths) having been reduced to that one smooth in $t$. Passing to
the resulting so-called $(3\!$ $\!+\!$ $\!1)\!$ Regge calculus had been undertaken in
a number of works \cite{ColWil1} - \cite{Tuc}. Usually, the authors of the cited
papers tried to define in some way the discrete analogs of the variables $p_A$, $q_A$
and of the constraints $\Phi_{\alpha}(p,q)$, the largest attention having been paid to
providing the algebra of Poisson brackets being maximally close to that in the
continuum GR. If we stick to the strategy requiring to deal with a realizable
Riemannian manifold at each stage, the $(3\!$ $\!+\!$ $\!1)\!$ Regge calculus follows
as the limiting case of the 4-dimensional Regge calculus while sizes of the
4-simplices tend to zero in some direction taken as direction of time. This limiting
procedure had been studied in the papers \cite{ColWil1,ColWil2,Bre1,Bre2}, although
not all the degrees of freedom could be taken into account in these works. The reason
is in the singular nature of description of Regge manifold with the help of the
linklengths when the scale of sizes along any direction tends to zero. As an example,
one can imagine a triangle one of the edges of which is infinitesimal; then
infinitesimal variations of the two other (finite) edge lengths lead to finite
variations of the angles. As a result, description in terms of only the lengths had
left some (singular) degrees of freedom missing, and therefore not all the discrete
counterparts of the constraints $\Phi_{\alpha}(p,q)$ could be found.

\medskip

{\bf 2.The problem of constructing the quantum measure in Regge calculus}

\medskip

The singular nature of passing to the continuous time is thus connected with using
only the lengths as fundamental set of variables in Regge calculus. While we are
studying the quantum measure on the {\it completely discrete} Regge manifold, this
circumstance is not of importance for us. However the issue concept of quantum theory
which might be used to construct quantum measure is canonical quantization. The latter
is defined just in the continuous time. Therefore the quantum measure of interest
should be defined from the requirement that it would tend in some sense to the
canonical quantization measure (Feynman path integral) whenever continuum limit is
taken along any of the coordinates, the coordinate chosen playing the role of time. In
other words, the continuous time limit serves as probing tool for defining the quantum
measure in the completely discrete Regge calculus.

The way to avoid singularities in the continuous time limit is in extending the set of
variables via adding the new ones having the sense of angles and considered as
independent variables. Such the variables are the finite rotation matrices which are
the discrete analogs of the connections in the continuum GR.

\medskip

{\bf 3.Representation of the Regge calculus in terms of finite rotation matrices as
independent variables}

\medskip

The situation considered is analogous to that one occurred when recasting the Einstein
action (\ref{S-Einst}) in the Hilbert-Palatini form,
\begin{equation}                                                                    
\label{S-HilPal} {1\over 2}\int{R\sqrt{g}{\rm d}^4x} \Leftarrow {1\over
8}\int{\epsilon_{abcd}\epsilon^{\lambda\mu\nu\rho}e^a_{\lambda}e^b_{\mu}
[\partial_{\nu}+\omega_{\nu},\partial_{\rho}+\omega_{\rho}]^{cd}{\rm d}^4x},
\end{equation}

\noindent where the tetrad $e^a_{\lambda}$ and connection $\omega^{ab}_{\lambda}$ =
$-\omega^{ba}_{\lambda}$ are independent variables, the equation (\ref{S-HilPal})
being reduced to (\ref{S-Einst}) in terms of $g_{\lambda\mu}$ =
$e^a_{\lambda}e_{a\mu}$ if we substitute for $\omega^{ab}_{\lambda}$ solution of the
equations of motion for these variables in terms of $e^a_{\lambda}$. The Latin indices
$a$, $b$, $c$, ... are the vector ones with respect to the local Euclidean frames
which are introduced at the each point $x$. The Regge calculus analog of the
representation (\ref{S-HilPal}) follows if the local Euclidean frames are introduced
in all the 4-simplices. Then the analogs of the connection are defined on the
3-simplices $\sigma^3$ and are the matrices $\Omega_{\sigma^3}$ connecting the frames
of the pairs of the 4-simplices $\sigma^4$ sharing the 3-faces $\sigma^3$. These
matrices are the finite SO(4) rotations in the Euclidean case (or SO(3,1) rotations in
the Lorentzian case) in contrast with the continuum connections
$\omega^{ab}_{\lambda}$ which are the elements of the Lee algebra so(4)(so(3,1)) of
this group. This definition includes pointing out the direction in which the
connection $\Omega_{\sigma^3}$ acts (and, correspondingly, the opposite direction, in
which the $\Omega^{-1}_{\sigma^3}$ = $\bar{\Omega}_{\sigma^3}$ acts), that is, the
connections $\Omega$ are defined on the {\it oriented} 3-simplices $\sigma^3$.

Also we can define the curvature matrix $R_{\sigma^2}$ on each the 2-simplex
$\sigma^2$ as the product of the connections $\Omega^{\pm 1}_{\sigma^3}$ on the
3-simplices $\sigma^3$ sharing $\sigma^2$ which act in certain direction along the
contour enclosing $\sigma^2$ once and contained in these 3-simplices. The matrix
$R_{\sigma^2}$ should be rotation around $\sigma^2$ by an angle $\alpha_{\sigma^2}$.
Besides the direction along the contour it is necessary to specify the 4-simplex
$\sigma^4$ $\supset$ $\sigma^2$ in which the contour begins and is ended, that is, the
simplex in the local Euclidean frame of which we define
\begin{equation}                                                                    
\label{R-Omega} R_{\sigma^2} = \prod_{\sigma^3\supset\sigma^2}{\Omega^{\pm
1}_{\sigma^3}}.
\end{equation}

\noindent The discrete analogs of the connection and curvature had been considered by
Bander \cite{Ban1,Ban2,Ban3} as functions of the lengths. Our approach is based on
treatment of the connections as independent variables and using a representation of
the Regge calculus action (\ref{S-Regge}) analogous to the Hilbert-Palatini form of
the Einstein action (\ref{S-HilPal}). To write out this representation let us define
the dual bivector of the triangle $\sigma^2$ in terms of the vectors of its edges
$l^a_1$, $l^a_2$ given in some 4-simplex containing $\sigma^2$,
\begin{equation}                                                                    
\label{v=ll} v_{\sigma^2ab} = {1\over 2}\epsilon_{abcd}l^c_1l^d_2.
\end{equation}

\noindent Then the discrete analog of the expression (\ref{S-HilPal}) as suggested in
our work \cite{Kha01} reads
\begin{equation}                                                                    
\label{S-RegCon}%
S(v,\Omega) = \sum_{\sigma^2}{|v_{\sigma^2}|\arcsin{v_{\sigma^2}\circ
R_{\sigma^2}(\Omega)\over |v_{\sigma^2}|}}
\end{equation}

\noindent where we have defined $A\circ B$ = ${1\over 2}A^{ab}B_{ab}$, $|A|$ =
$(A\circ A)^{1/2}$ for the two tensors $A$, $B$, in particular, $|v_{\sigma^2}|$ =
$|\sigma^2|$ is the area of the triangle. It is important that $v_{\sigma^2}$ and
$R_{\sigma^2}$ in (\ref{S-RegCon}) be defined in the same 4-simplex containing
$\sigma^2$. As we can show, when substituting as $\Omega_{\sigma^3}$ the genuine
rotations connecting the neighbouring local frames as functions of the genuine Regge
lengths into the equations of motion for $\Omega_{\sigma^3}$ with the action
(\ref{S-RegCon}) we get the closure condition for the surface of the 3-simplex
$\sigma^3$ (vanishing the sum of the bivectors of its 2-faces) written in the frame of
one of the 4-simplices containing $\sigma^3$, that is, the identity. This means that
(\ref{S-RegCon}) is the exact representation for (\ref{S-Regge}). At the same time the
paper \cite{CasDadMag} has appeared which suggests a representation which reminds
(\ref{S-RegCon}) but differs from it by replacing $\arcsin{(\cdot)}$ by $(\cdot)$.
This can be considered as representation for an approximate (at small angle defects)
Regge calculus with action
$\sum\limits_{\sigma^2}{|\sigma^2|\sin{\alpha_{\sigma^2}}}$. Here we should point out
remarkable feature namely of the exact Regge calculus since it is namely for the
action (\ref{S-RegCon}) the equations of motion for $\Omega_{\sigma^3}$ are shown by
us to hold {\it exactly} as closure condition for the surface of $\sigma^3$. In other
words, representation for an approximate Regge action inevitably proves to be, in
turn, approximate one.

\medskip

{\bf 4.Naturalness of arising area tensor Regge calculus}

\medskip

In the representation in terms of rotation matrices it is possible to pass to the
continuous time and develop the canonical formalism in Regge calculus \cite{Kha02}
which turns out to possess the second class constraints (that is, noncommuting ones).
As a result, Feynman path integral contains the determinant of the Poisson brackets of
the second class constraints as a factor which is singular when approaching the flat
geometry. The matter is that the Regge geometry usually changes at arbitrary
variations of the edge lengths with exception of the flat case in which these
variations are the symmetry transformations. In other words, division of the
constraints into those ones of the first and the second class is being changed in the
flat case. An exception is the 3-dimensional case. Due to the local triviality of the
3-dimensional gravity all the dynamical constraints are of the first class, and
therefore the path integral takes a simple form. In this case the problem of
constructing the discrete quantum measure as formulated in the subsection 2 can be
solved yielding a simple form for this measure \cite{Kha11}.

The conditions of the subsection 2 on the discrete measure are rather restrictive
ones, and existence of the solution is not evident. Singularity of the path integral
in 4 dimensions in the vicinity of the flat background is by itself not an obstacle
for existence of such the solution; crucial is the occurrence of the above determinant
factor in the path integral. This factor depends on the variables which are lattice
artefacts connected with {\it specific} coordinate along which the continuum limit is
taken, and it can not be obtained from some {\it universal} expression by taking the
continuous time limit.

Let us try to modify the 4-dimensional Regge calculus so that it would remind the
3-dimensional case as far as canonical structure is concerned. The 3-dimensional Regge
calculus in the representation analogous to (\ref{S-RegCon}) has the edge vectors
$\bl_{\sigma^1}$ instead of the area tensors $v_{\sigma^2}$. The edge vectors are the
independent variables thus ensuring the local triviality of the 3-dimensional gravity.
Contrary to this, area tensors are not independent. For example, tensors of the two
triangles $\sigma^2_1$, $\sigma^2_2$ sharing an edge satisfy the relation
\begin{equation}                                                                    
\label{v*v}%
\epsilon_{abcd}v^{ab}_{\sigma^2_1}v^{cd}_{\sigma^2_2} = 0.
\end{equation}

\noindent The idea is to construct quantum measure first for the system with formally
independent area tensors. That is, originally we concentrate on quantization of the
dynamics while kinematical relations of the type (\ref{v*v}) are taken into account at
the second stage.

Area tensor Regge calculus admits solution to the problem of constructing the discrete
quantum measure, the latter taking a simple form \cite{Kha12}. Consider the Euclidean
case. The Einstein action is not bounded from below, therefore the Euclidean path
integral itself requires careful definition. In particular, the result of \cite{Kha12}
for vacuum expectations of the functions of our field variables $v$, $\Omega$ can be
written with the help of the integration over imaginary areas with the help of the
formal replacement of the tensors of a certain subset of areas $\pi$ over which
integration in the path integral is to be performed, $$\pi \rightarrow -i\pi,$$ in the
form
\begin{eqnarray}                                                                   
\label{VEV2}%
<\Psi (\{\pi\},\{\Omega\})> & = & \int{\Psi (-i\{\pi\}, \{\Omega\})\exp{\left (-\!
\sum_{\stackrel{t-{\rm like}}{\sigma^2}}{\tau _{\sigma^2}\circ
R_{\sigma^2}(\Omega)}\right )}}\nonumber\\
 & & \hspace{-20mm} \cdot \exp{\left (i
\!\sum_{\stackrel{\stackrel{\rm not}{t-{\rm like}}}{\sigma^2}} {\pi_{\sigma^2}\circ
R_{\sigma^2}(\Omega)}\right )}\prod_{\stackrel{\stackrel{\rm
 not}{t-{\rm like}}}{\sigma^2}}{\rm d}^6
\pi_{\sigma^2}\prod_{\sigma^3}{{\cal D}\Omega_{\sigma^3}} \nonumber\\ & \equiv &
\int{\Psi (-i\{\pi\},\{\Omega\}){\rm d} \mu_{\rm area}(-i\{\pi\},\{\Omega\})},
\end{eqnarray}

\noindent where we define $A\circ B$ = ${1\over 2}A^{ab}B_{ab}$ for the two tensors
$A$, $B$. The equation implies attributing a certain structure to our Regge lattice
which suggests constructing it of leaves being themselves the 3-dimensional Regge
geometries of the same structure. The leaves are labelled by the values of some
coordinate $t$. The corresponding vertices in the neighbouring leaves are connected by
the {\it $t$-like} edges, and, in addition, there are the {\it diagonal} edges
connecting a vertex with the neighbours of the corresponding vertex in the
neighbouring leaf. Then it is natural to define the $t$-like simplices and the leaf
simplices as the simplices either containing a $t$-like edge or completely contained
in the leaf, respectively, and also the diagonal simplices as all others. Then
$\tau_{\sigma^2}$ is $v_{\sigma^2}$ when $\sigma^2$ is $t$-like while $\pi_{\sigma^2}$
is $v_{\sigma^2}$ when $\sigma^2$ is not $t$-like, that is, is the leaf or diagonal
simplex. As far as considered here area tensor Regge calculus is concerned (the area
tensors are independent), the $\pi_{\sigma^2}$ can be chosen as dynamical variables,
then the $\tau_{\sigma^2}$ should be considered as parameters.

The equation (\ref{VEV2}) is in many aspects similar to the intuitively expected one
for the quantum measure. In particular, the expected from symmetry considerations
invariant (Haar) measure on SO(4) ${\cal D}\Omega$ arises in the formal path integral
expression corresponding in the continuous time limit to the canonical quantization
with the kinetic term $\pi_{\sigma^2}\circ \bar{\Omega}_{\sigma^2}
\dot{\Omega}_{\sigma^2}$ in the Lagrangian (the connection variables $\Omega$ in the
continuous time limit naturally correspond not to the tetrahedra $\sigma^3$ but to the
triangles $\sigma^2$).

Specific features of the quantum measure are, first, the absence of the inverse
trigonometric functions $\arcsin$ in the exponential, whereas the Regge action
($\ref{S-RegCon}$) contains such functions. This is connected with using the canonical
quantization at the intermediate stage of derivation: in gravity this quantization is
completely defined by the constraints, the latter being equivalent to those ones
without $\arcsin$ (in some sense on-shell).

Second, there are no integrations over some subset of area tensors, $\tau_{\sigma^2}$,
thereby the symmetry between the different triangles turns out to be incomplete.
However, this violation of symmetry can be considered as spontaneous one when some
a'priori arbitrary direction coordinatised in (\ref{VEV2}) by $t$ turns out to be
singled out. The curvature matrices $R(\Omega)$ on all but $t$-like triangles can be
chosen as independent variables, then such matrices on the $t$-like triangles are by
means of the Bianchi identities the functions of these variables. Integrations over
{\it all} the area tensors would lead to singularities of the type of $[\delta (R$
$\!-\!$ $\bar{R})]^2$.

This specific feature of the discrete quantum measure, incomplete symmetry with
respect to the different coordinate directions, complies with the above conditions of
subsection 2 in the following way. In the continuous limit along some coordinate $x$
(which may not coincide with $t$) the absence of integrations over the tensors of the
$t$-like triangles will mean some of the simplest kinds of gauge fixing in the
limiting measure, namely, fixing the tensors of some subset of the triangles
\cite{Kha12}.

With taking into account the properties of the invariant Haar measure and with the
negligibly small values of $\tau_{\sigma^2}$ we get factorisation of the quantum
measure obtained into the "elementary" measures on the separate areas (this just
corresponds to the local triviality of the theory) of the form
\begin{equation}                                                                   
\label{separate} \exp{(i\pi\circ R)}{\rm d}^6\pi{\cal D}R.
\end{equation}

\noindent In turn, use the group property SO(4) = SU(2) $\times$ SU(2) to split the
variables ($\pi$ and generator of $R$) into the self- and antiselfdual parts, in
particular, $\pi$ is mapped into the two 3-vectors $\,^{+}\!\bpi$, $\,^{-}\!\bpi$ in
the adjoint representation SO(3). As a result, the measure (\ref{separate}) is the
product of the two measures each of which act in 3-dimensional configuration space of
area vectors,
\begin{equation}                                                                   
\label{+d-mu-d-mu}%
\exp(i\,^+\!\pi\circ\,^+\! R)\d^3\pbpi{\cal D}\,^+\! R\cdot \exp(i\,^-\!\pi\circ\,^-\!
R)\d^3\mbpi{\cal D}\,^-\! R.
\end{equation}

\noindent As a result, expression for the expectation of any function on the triangle
reads
\begin{eqnarray}                                                                   
\label{Euclidean_measure}%
<f(\pi)> & = & \int{f (-i\pi ){\rm d}^6\pi\int{e^{\textstyle i\pi\circ R}{\cal D}R}}
\nonumber\\ & = & \int{f (\pi ) {\nu_2 (|\pbpi |)\over |\pbpi |^2}{\nu (|\mbpi |)\over
|\mbpi |^2} {\d^3\pbpi\over 4\pi} {\d^3\mbpi\over 4\pi}}, \\ & & \nu(l)={l\over\pi}
\int\limits_{0}^{\pi}{{\rm d}\varphi\over\sin^2{\!\varphi}}\,{e^{\textstyle
-l/\sin{\varphi}}}. \nonumber
\end{eqnarray}

\noindent In particular, expectations of powers of area squared,
\begin{equation}                                                                   
|\pi|^2 = \pi\circ\pi = {1\over 2}(\,^{+}\!\bpi)^2 + {1\over 2}(\,^{-}\!\bpi)^2,
\end{equation}

\noindent and of the dual product,
\begin{equation}                                                                   
\label{dual} \pi*\pi={1\over 2}(\,^{+}\!\bpi)^2 - {1\over 2}(\,^{-}\!\bpi)^2,
\end{equation}

\noindent easily follow by means of averaging the powers of $\pmbpi$,
\begin{equation}                                                                   
\label{<bpi>}%
<(\pmbpi)^{2k}> = {4^{-k}(2k+1)!(2k)!\over k!^2}.
\end{equation}

\medskip

{\bf 5.Reducing to genuine Regge calculus case}

\medskip

Thus, we have come to the finite nonzero area expectation values in area tensor Regge
calculus. However, we need the {\it length} expectations in the genuine ordinary Regge
calculus. The ordinary Regge calculus follows upon imposing unambiguity condi\-tions
on the lengths computed in the different 4-simplices. These condi\-tions are
equivalent to the continuity condi\-tions for the induced on 3-faces metric. In the
configuration space of area tensor Regge calculus these conditions single out some
hypersurface $\Gamma_{\rm Regge}$. The quantum measure can be considered as a linear
functional $\mu_{\rm area}(\Psi)$ on the space of functionals $\Psi (\{v\})$ on the
configuration space (for our purposes here it is sufficient to restrict ourselves to
the functional dependence on the area tensors $\{v\}$; the dependence on the
connections is unimportant). The physical assumption is that we can consider ordinary
Regge calculus as a kind of the state of the more general system with independent area
tensors. This state is described by the following functional,
\begin{equation}                                                                   
\Psi (\{v\}) = \psi (\{v\})\delta_{\rm Regge}(\{v\}),
\end{equation}

\noindent where $\delta_{\rm Regge}(\{v\})$ is the (many-dimensional)
$\delta$-function with support on $\Gamma_{\rm Regge}$. The derivatives of
$\delta_{\rm Regge}$ have the same support, but these violate positivity in our
subsequent construction. To be more precise, delta-function is distribution, not
function, but can be treated as function being regularised. If the measure on such the
functionals exists in the limit when regularisation is removed, this allows to define
the quantum measure on $\Gamma_{\rm Regge}$,
\begin{equation}                                                                   
\label{mu-proj}%
\mu_{\rm Regge}(\cdot) = \mu_{\rm area}(\delta _{\rm Regge}(\{v\})~\cdot).
\end{equation}

Uniqueness of the construction of $\delta_{\rm Regge}$ follows under quite natural
assumption of the minimum of lattice artefacts. Let the system be described by the
metric $g_{\lambda\mu}$ constant in each of the two 4-simplices $\sigma^4_1$,
$\sigma^4_2$ separated by the 3-face $\sigma^3$ = $\sigma^4_1\cap \sigma^4_2$ formed
by three 4-vectors $\iota^\lambda_a$. These vectors also define the metric induced on
the 3-face, $g^{\|}_{ab}$ = $\iota^\lambda_a\iota^\mu_bg_{\lambda\mu}$. The continuity
condition for the induced metric is taken into account by the $\delta$-function of the
induced metric variation,
\begin{equation}                                                                   
\label{delta-g} \Delta_{\sigma^3}g^{\|}_{ab}\stackrel{\rm def}{=}
g^{\|}_{ab}(\sigma^4_1)-g^{\|}_{ab}(\sigma^4_2).
\end{equation}

\noindent As for the $\delta_{\rm Regge}$, it is of course defined up to a factor
which is arbitrary function nonvanishing at nondegenerate field configurations. In the
spirit of just mentioned principle of minimizing the lattice artefacts it is natural
to choose this factor in such the way that the resulting $\delta$-function factor
would depend only on hyperplane defined by the 3-face but not on the form of this
face, that is, would be invariant with respect to arbitrary nondegenerate
transformations $\iota^\lambda_a$ $\mapsto$ $m^b_a\iota^\lambda_b$. To provide this,
the $\delta$-function should be multiplied by the determinant $g^{\|}_{ab}$ squared.
This gives
\begin{equation}                                                                   
\label{inv-delta}%
[{\rm det}(\iota^\lambda_a\iota^\mu_bg_{\lambda\mu})]^2
\delta^6(\iota^\lambda_a\iota^\mu_b \Delta_{\sigma^3}g_{\lambda\mu}) = V^4_{\sigma^3}
\delta^6(\Delta_{\sigma^3}S_{\sigma^3}).
\end{equation}

\noindent Here $S_{\sigma^3}$ is the set of the 6 edge lengths squared of the 3-face
$\sigma^3$, $V_{\sigma^3}$ is the volume of the face.

Further, the product of the factors (\ref{inv-delta}) over all the 3-faces should be
taken. As a result, we have for each edge the products of the $\delta$-functions of
the variations of its length between the 4-simplices taken along closed contours,
$\delta (s_1-s_2)\delta (s_2-s_3)\dots\delta (s_N-s_1)$ containing singularity of the
type of the $\delta$-function squared. In other words, the conditions equating
(\ref{delta-g}) to zero on the different 3-faces are not independent. The more
detailed consideration allows to cancel this singularity in a way symmetrical with
respect to the different 4-simplices (thus extracting irreducible conditions), the
resulting $\delta$-function factor remaining invariant with respect to arbitrary
deformations of the faces of different dimensions keeping each face in the fixed plane
spanned by it \cite{Kha14}.

Qualitatively, it is important that our $\delta$-function factor (see, for example,
the simplest version (\ref{inv-delta})) automatically turns out to be invariant with
respect to overall length scaling. Remind, however, that dynamical variables to be
averaged are $\pi_{\sigma^2}$ but not $\tau_{\sigma^2}$. The more detailed analysis
shows that upon fixing the scale of the tensors $\tau_{\sigma^2}$ at the level
$\varepsilon\!$ $\!\ll\!\!$ 1 the $\delta$-function factor is invariant also with
respect to overall scaling of only the dynamical variables $\pi_{\sigma^2}$. This
leads to the finite nonzero length expectation values in the ordinary Regge calculus
as far as the area expectation values in area tensor Regge calculus are finite and
nonzero \cite{Kha15}.

Strictly speaking, when passing from area tensor Regge calculus to the ordinary Regge
calculus we need first to impose the conditions ensuring that tensors of the 2-faces
in the given 4-simplex define a metric in this simplex. These conditions of the type
of (\ref{v*v}) can be easily written in general form. Let a vertex of the given
4-simplex be the coordinate origin and the edges emitted from it be the coordinate
lines $\lambda$, $\mu$, $\nu$, $\rho$, \dots = 1, 2, 3, 4. Then the (ordered) pair
$\lambda\mu$ means the (oriented) triangle formed by the edges $\lambda$, $\mu$. The
conditions of interest take the form
\begin{equation}                                                                   
\label{tetrad} \epsilon_{abcd}v^{ab}_{\lambda\mu}v^{cd}_{\nu\rho}
\sim\epsilon_{\lambda\mu\nu\rho}.                        
\end{equation}

\noindent The 20 equations (\ref{tetrad}) define the 16-dimensional surface $\gamma
(\sigma^4)$ in the 36-dimensional configura\-tion space of the six antisymmetric
tensors $v^{ab} _{\lambda\mu}$\footnote{There are also the linear constraints of the
type $\sum{\pm v}$ = 0 providing closing surfaces of the 3-faces of our 4-simplex.
These are assumed to be already resolved.}. The factor of interest in quantum measure
is the product of the $\delta$-functions with support on $\gamma (\sigma^4)$ over all
the 4-simplices $\sigma^4$. The covariant form of the constraints (\ref{tetrad}) with
respect to the world index means that these $\delta$-functions are the scalar
densities of a certain weight with respect to the world index, that is, the scalars up
to powers of the volume of the 4-simplex $V_{\sigma^4}$. Therefore introducing the
factors of the type $V_{\sigma^4}^{\eta}$ we get the scalar at some parameter $\eta$.
Namely, the product of the factors
\begin{equation}                                                                   
\label{delta-metric}%
\prod_{\sigma^4}{\int{V_{\sigma^4}
^{\eta}\delta^{21}(\epsilon_{abcd}v^{ab}_{\lambda\mu | \sigma^4}v^{cd}_{\nu\rho
|\sigma^4} - V_{\sigma^4}\epsilon _{\lambda\mu\nu\rho})\,{\rm d}V_{\sigma^4}}}
\end{equation}

\noindent at $\eta$ = 20 is the scale invariant value as required by the principle of
the minimum of the lattice artefacts (that is, the factor of interest should not
depend on the size of the 4-simplex). As a result, the conclusion made in the previous
paragraph that the length expectation values in ordinary Regge calculus are finite and
nonzero as soon as the area expectation values in area tensor Regge calculus are
finite and nonzero remains valid \cite{Kha15}.

\medskip

{\bf 6.Conclusion}

\medskip

Thus, our approach to quantization of Regge calculus "from the first principles"
includes the following steps and conditions.

\noindent 1. Constructing the quantum measure which is reduced to the Feynman path
integral corres\-pon\-ding to the canonical quantization in the continuous time limit
irrespectively of what coordinate is taken as time.

\noindent 2. Using the exact representation of the Regge action in terms of the
rotation matrices as independent variables.

\noindent 3. Extending the configuration space of the theory by considering area
tensors as independent variables (considering the so-called area tensor Regge
calculus).

\noindent 4. Reducing the quantum measure from area tensor Regge calculus to the
hypersurface corres\-pon\-ding to the ordinary Regge calculus under assumption of the
minimal lattice artefacts, that is, minimal dependence on the form and size of the
simplices.

\noindent As a result, the length expectations are found to be of the order of the
Plank scale $10^{-33}cm$. Were these values vanishing, this would mean that the
quantum measure is saturated by the arbitrarily small lengths, that is, by the smooth
Riemannian manifolds, and we would return to the continuum GR. Here the remarkable
property of the Regge calculus is displayed: it is minisuperspace GR theory, that is,
it is exact GR for a class of certain (piecewise flat) spacetimes. Therefore Regge
calculus in quantum theory does not mean complete exclusion of the continuum GR
(rather it contains the continuum GR as the limiting point), but rather presents an
alternative way of description of the system with the help of the triangulation
lengths. Our result, the nonzero length expectations, means adequacy of namely such
the description. The GR becomes discrete at the Plank scale {\it dynamically}, that
is, as a result of competition between the different contributions into the functional
integral including also that one from the smooth manifolds.

I am grateful to I.B. Khriplovich for attention to the work and discussion. The
present work was supported in part by the Russian Foundation for Basic Research
through Grant No. 05-02-16627-a.

\end{document}